\documentclass{article}
\pdfoutput=1
\usepackage{authblk}
\usepackage[margin=1in]{geometry}
\usepackage[utf8]{inputenc}
\usepackage{graphicx}
\usepackage{amsmath}
\usepackage{amsfonts}
\usepackage{color, colortbl}
\usepackage[colorlinks = true,
            linkcolor = blue,
            urlcolor  = blue,
            citecolor = blue]{hyperref}

\usepackage[mathlines]{lineno}
\usepackage[round]{natbib}   

\newcommand{\tinyff}[1]{\scalebox{0.7}{{\normalsize #1}}}		
\usepackage{import}
\usepackage{lineno}

\usepackage[mathscr]{eucal}
\renewcommand{\O}{{\mathscr{O}}}
\newcommand{\obs}{obs}
\newcommand{\pred}{pred}
\renewcommand{\inf}{inf}

\usepackage{amssymb}
\usepackage{enumitem}
\usepackage{amsthm}
\newtheorem{Def}{Definition}[section]
\newtheorem{Lemma}[Def]{Lemma}
\newtheorem{Prp}[Def]{Proposition}
\newtheorem{Thm}[Def]{Theorem}
\newcommand{\Proof}{\begin{proof}}
\newcommand{\QED}{\end{proof} \noindent}
\newcommand{\Obar}{\overline \O}

\usepackage[dvipsnames]{xcolor}

\title{Falsification and consciousness}

\author[1]{
Johannes Kleiner\thanks{Johannes.Kleiner@lmu.de}}
\author[2]{Erik Hoel\thanks{corresponding author: erik.hoel@tufts.edu}}
\affil[1]{Munich Center for Mathematical Philosophy, Ludwig Maximilian University of Munich}
\affil[2]{Allen Discovery Center, Tufts University, Medford, MA, USA}

\begin{document}

\maketitle

\begin{abstract}
The search for a scientific theory of consciousness should result in theories that are falsifiable. However, here we show that falsification is especially problematic for theories of consciousness. We formally describe the standard experimental setup for testing these theories. Based on a theory's application to some physical system, such as the brain, testing requires comparing a theory's predicted experience (given some internal observables of the system like brain imaging data) with an inferred experience (using report or behavior). If there is a mismatch between inference and prediction, a theory is falsified. We show that if inference and prediction are independent, it follows that any minimally informative theory of consciousness is automatically falsified. This is deeply problematic since the field's reliance on report or behavior to infer conscious experiences implies such independence, so this fragility affects many contemporary theories of consciousness. Furthermore, we show that if inference and prediction are strictly dependent, it follows that a theory is unfalsifiable. This affects theories which claim consciousness to be determined by report or behavior. Finally, we explore possible ways out of this dilemma. 
\end{abstract}

\section{\label{sec:introduction}Introduction}

Successful scientific fields move from exploratory studies and observations to the point where theories are proposed that can offer precise predictions. Within neuroscience the attempt to understand consciousness has moved out of the exploratory stage and there are now a number of theories of consciousness capable of predictions that have been advanced by various authors \citep{koch2016neural}. 

At this point in the field's development falsification has become relevant. In general, scientific theories should strive to make testable predictions \citep{popper2005logic}. In the search for a scientific theory of consciousness, falsifiability must be considered explicitly as it is commonly assumed that consciousness itself cannot be directly observed, instead it can only be inferred based off of report or behavior.

Contemporary neuroscientific theories of consciousness first began to be proposed in the early 1990s \citep{crick1994astonishing}. Some have been based directly on neurophysiological correlates, such as proposing that consciousness is associated with neurons firing at a particular frequency \citep{crick1990towards} or activity in some particular area of the brain like the claustrum \citep{crick2005function}. Other theories have focused more on the dynamics of neural processing, such as the degree of recurrent neural connectivity \citep{lamme2006towards}. Others yet have focused on the ``global workspace'' of the brain, based on how signals are propagated across different brain regions \citep{baars1997theatre}. Specifically, Global Neuronal Workspace theory claims that consciousness is the result of an ``avalanche'' or ``ignition'' of widespread neural activity created by an interconnected but dispersed network of neurons with long-range connections \citep{sergent2004neural}. 

Another avenue of research strives to derive a theory of consciousness from analysis of phenomenal experience. The most promising example thereof is Integrated Information Theory \citep{tononi2004information, tononi2008consciousness, oizumi2014phenomenology}. Historically, Integrated Information Theory is the first well-formalized theory of consciousness. It was the first (and arguably may still be the lone) theory that makes precise quantitative predictions about both the contents and level of consciousness \citep{tononi2004information}. Specifically, the theory takes the form of a function, the input of which is data derived from some physical system's internal observables, while the output of this function are predictions about the contents of consciousness (represented mathematically as an element of an experience space) and the level of consciousness (represented by a scalar value $\Phi$).

Both Global Neuronal Workspace (GNW) and Integrated Information Theory (IIT) have gained widespread popularity, sparked general interest in consciousness, and have led to dozens if not hundreds of new empirical studies \citep{massimini2005breakdown, del2007brain, dehaene2011experimental, gosseries2014measuring, wenzel2019reduced}. Indeed, there are already significant resources being spent attempting to falsify either GNW or IIT in the form of a global effort pre-registering predictions from the two theories so that testing can be conducted in controlled circumstances by researchers across the world \citep{TempletonProject,reardon2019rival}. We therefore often refer to both GNW and IIT as exemplar theories within consciousness research and show how our results apply to both. However, our results and reasoning apply to most contemporary theories, e.g.  \citep{lau2011empirical, chang2019information}, particularly in their ideal forms. Note that we refer to both ``theories'' of consciousness and also ``models'' of consciousness, and use these interchangeably \citep{seth2007models}.

Due to IIT's level of formalization as a theory, it has triggered the most in-depth responses, expansions, and criticisms \citep{cerullo2015problem, bayne2018axiomatic, mediano2019measuring, kleiner2020mathematical} since well-formalized theories are much easier to criticize than non-formalized theories. Recently one criticism levied against IIT was based on how the theory predicts feedfoward neural networks have zero $\Phi$ and recurrent neural networks have non-zero $\Phi$. Since a given recurrent neural network can be ``unfolded'' into a feedfoward one while preserving its output function, this has been argued to render IIT outside the realm of science \citep{doerig2019unfolding}. Replies have criticised the assumptions which underlie the derivation of this argument \citep{kleiner2019empirical, tsuchiya2019reply}.

Here we frame and expand concerns around testing and falsification of theories by examining a more general question: what are the conditions under which theories of consciousness (beyond IIT alone) can be falsified? We outline a parsimonious description of theory testing with minimal assumptions based on first principles. In this agnostic setup falsifying a theory of consciousness is the result of finding a mismatch between the inferred contents of consciousness (usually based on report or behavior) and the contents of consciousness as predicted by the theory (based on the internal observables of the system under question).

This mismatch between prediction and inference is critical for an empirically meaningful scientific agenda, because a theory's prediction of the state and content of consciousness on its own cannot be assessed. For instance, imagine a theory that predicts (based on internal observables like brain dynamics) that a subject is seeing an image of a cat. Without any reference to report or outside information, there can be no falsification of this theory, since it cannot be assessed whether the subject was actually seeing a ``dog'' rather than ``cat.'' Falsifying a theory of consciousness is based on finding such mismatches between reported experiences and predictions.

In the following work, we formalize this by describing the prototypical experimental setup for testing a theory of consciousness. We come to a surprising conclusion: a widespread experimental assumption implies that most contemporary theories of consciousness are already falsified.

The assumption in question is the \textit{independence} of an experimenter's inferences about consciousness from a theory's predictions. To demonstrate the problems this independence creates for contemporary theories, we introduce a ``substitution argument.'' This argument is based on the fact that many systems are equivalent in their reports (e.g., their outputs are identical for the same inputs) and yet their internal observables may differ greatly. This argument constitutes both a generalization and correction of the ``unfolding argument'' against IIT presented in~\cite{doerig2019unfolding}. Examples of such substitutions may involve substituting a brain with a Turing machine or a cellular automaton since both types of systems are capable of universal computation \citep{turing1937computable, wolfram1984cellular} and hence may emulate the brain's responses, or replacing a deep neural network with a single-layer neural network, since both types of networks can approximate any given function \citep{hornik1989multilayer, schafer2006recurrent}.

Crucially, our results do not imply that falsifications are impossible. Rather, they show that the independence assumption implies that whenever there is an experiment where a theory's predictions based on internal observables and a system's reports agree, there exists also an actual physical system that falsifies the theory. One consequence is that the ``unfolding argument'' concerning IIT \citep{doerig2019unfolding} is merely a small subset of a much larger issue that affects all contemporary theories which seek to make predictions about experience off of internal observables. Our conclusion shows that if independence holds, all such theories come falsified \textit{a priori}. Thus, instead of putting the blame of this problem on individual theories of consciousness, we show that it is due to issues of falsification in the scientific study of consciousness, particularly the field's contemporary usage of report or behavior to infer conscious experiences.

A simple response to avoid this problem is to claim that report and inference are not independent. This is the case, e.g., in behaviorist theories of consciousness, but arguably also in Global Workspace Theory \citep{baars2005global}, the ``attention schema'' theory of consciousness \citep{graziano2015attention} or ``fame in the brain'' \citep{dennett1993consciousness} proposals. We study this answer in detail and find that making a theory's predictions and an experimenter's inferences \textit{strictly dependent} leads to pathological unfalsifiability.

Our results show that if independence of prediction and inference holds true, as in contemporary cases where report about experiences is relied upon, it is likely that no current theory of consciousness is correct. Alternatively, if the assumption of independence is rejected, theories rapidly become unfalsifiable. While this dilemma may seem like a highly negative conclusion, we take it to show that our understanding of testing theories of consciousness may need to change to deal with these issues.

\section{Formal description of testing theories}\label{formal_description_theory}

Here we provide a formal framework for experimentally testing a particular class of theories of consciousness. The class we consider makes \textit{predictions} about the \textit{conscious experience} of \textit{physical systems} based on \textit{observations or measurements}. This class describes many contemporary theories, including leading theories such as Integrated Information Theory \citep{oizumi2014phenomenology}, Global Neuronal Workspace Theory \citep{dehaene2004neural}, Predictive Processing (when applied to account for conscious experience \citep{dolkega2020fame,clark2019consciousness,seth2014predictive,hobson2014consciousness,hohwy2012attention}) or Higher Order Thought Theory \citep{rosenthal2002many}. These theories may be motivated in different ways, or contain different formal structures, such as for example the ones of category theory \citep{tsuchiya2016using}. In some cases, contemporary theories in this class may lack the specificity to actually make precise predictions in their current form. Therefore, the formalisms we introduce may sometimes describe a more advanced form of a theory, one that can actually make predictions. 

In the following section, we introduce the necessary terms to define how to falsify this class of theories: how measurement of a physical system's observables results in datasets (Section \ref{sec:Experiments}), how a theory makes use of those datasets to offer predictions about consciousness (Section \ref{sec:Predictions}), how an experimenter makes inferences about a physical system's experiences (Section \ref{sec:Inferences}), and finally how falsification of a theory occurs when there is a mismatch between a theory's prediction and an experimenter's inference (Section \ref{sec:Falsification}). In Section \ref{sec:Summary} we give a summary of the introduced terms. In subsequent sections we explore the consequences of this setup, such as how all contemporary theories are already falsified if the data used by inferences and predictions are independent, and also how theories are unfalsifiable if this is changed to a strict form of dependency.

\subsection{Experiments}\label{sec:Experiments}

All experimental attempts to either falsify or confirm a member of the class of theories we consider begin by examining some particular physical system which has some specific physical configuration, state, or dynamics,~$p$. This physical system is part of a class $P$ of such systems which could have been realized, in principle, in the experiment. For example, in IIT, the class of systems $P$ may be some Markov chains, set of logic gates, or neurons in the brain, and every $p \in P$ denotes that system being in a particular state at some time $t$. On the other hand, for Global Neuronal Workspace, $P$ might comprise the set of long-range cortical connections that make up the global workspace of the brain, with $p$ being the activity of that global workspace at that time. 

Testing a physical system necessitates experiments or observations. For instance, neuroimaging tools like fMRI or EEG have to be used in order to obtain information about the brain. This information is used to create datasets such as functional networks, wiring diagrams, models, or transition probability matrices. To formalize this process, we denote by $\O$ all possible datasets that can result from observations of $P$. Each $o \in \O$ is one particular dataset, the result of carrying out some set of measurements on $p$. We denote the datasets that can result from measurements on $p$ as $\obs(p)$. Formally:

\begin{equation}\label{eq:obs}
\obs: P \twoheadrightarrow \O \:,
\end{equation}
where $\obs$ is a \emph{correspondence}, which is a ``generalized function'' that allows more than one element in the image $\obs(p)$ (functions are a special case of correspondences). A correspondence is necessary because, for a given $p$, various possible datasets may arise, e.g., due to different measurement techniques such as fMRI vs. EEG, or due to the stochastic behaviour of the system, or due to varying experimental parameters. In the real world, data obtained from experiments may be incomplete or noisy, or neuroscientific findings difficult to reproduce \citep{gilmore2017progress}. Thus for every $p \in P$, there is a whole class of datasets which can result from the experiment.

Note that $\obs$ describes the experiment, the choice of observables, and all conditions during an experiment that generates the dataset $o$ necessary to apply the theory, which may differ from theory to theory, such as interventions in the case of IIT. In all realistic cases, the correspondence $\obs$ is likely quite complicated since it describes the whole experimental setup. For our argument it simply suffices that this mapping exists, even if it is not known in detail.

It is also worth noting here that all leading neuroscientific theories of consciousness, from IIT to GNW, assume that experiences are not observable or directly measurable when applying the theory to physical systems. That is, experiences themselves are never identified or used in $\obs$, but are rather inferred based on some dataset $o$ that contains report or other behavioural indicators.

Next we explore how the datasets in $\O$ are used to make predictions about the experience of a physical system.

\subsection{Predictions}\label{sec:Predictions}

A theory of consciousness makes predictions about the experience of some physical system in some configuration, state, or dynamics, $p$, based on some dataset $o$. To this end, a theory carries within its definition a set or space $E$ whose elements correspond to various different \emph{conscious experiences} a system could have. The interpretation of this set varies from theory to theory, ranging from descriptions of the level of conscious experience in early versions of IIT, descriptions of the level and content of conscious experience in contemporary IIT \citep{kleiner2020mathematical}, or the description only of whether a presented stimuli is experienced in GNW or HOT. We sometimes refer to elements $e$ of $E$ simply as  \emph{experiences}. 

Formally, this means that a prediction considers an experimental dataset $o \in O$ (determined by $obs$) and specifies an element of the experience space $E$. We denote this as $pred$, for ``prediction,'' which is a map from $\O$ to $E$. The details of how individual datasets are being used to make predictions again do not matter for the sake of our investigation. What matters is that a procedure exists, and this is captured by $pred$. However, we have to take into account that a single dataset $o \in O$ may not predict only one single experience. In general, $pred$ may only allow an experimenter to constrain experience of the system in that it only specifies a subset of all experiences a theory models. We denote this subset by $\pred(o)$. Thus, $\pred$ is also a correspondence
\[
    \pred: \O \twoheadrightarrow E \:.
\]

\begin{figure}[t]
\centering
\begingroup%
  \makeatletter%
  \providecommand\color[2][]{%
    \errmessage{(Inkscape) Color is used for the text in Inkscape, but the package 'color.sty' is not loaded}%
    \renewcommand\color[2][]{}%
  }%
  \providecommand\transparent[1]{%
    \errmessage{(Inkscape) Transparency is used (non-zero) for the text in Inkscape, but the package 'transparent.sty' is not loaded}%
    \renewcommand\transparent[1]{}%
  }%
  \providecommand\rotatebox[2]{#2}%
  \newcommand*\fsize{\dimexpr\f@size pt\relax}%
  \newcommand*\lineheight[1]{\fontsize{\fsize}{#1\fsize}\selectfont}%
  \ifx\svgwidth\undefined%
    \setlength{\unitlength}{225.66016875bp}%
    \ifx\svgscale\undefined%
      \relax%
    \else%
      \setlength{\unitlength}{\unitlength * \real{\svgscale}}%
    \fi%
  \else%
    \setlength{\unitlength}{\svgwidth}%
  \fi%
  \global\let\svgwidth\undefined%
  \global\let\svgscale\undefined%
  \makeatother%
  \begin{picture}(1,0.08170089)%
    \lineheight{1}%
    \setlength\tabcolsep{0pt}%
    \put(-0.00514982,0.00918666){\color[rgb]{0,0,0}\makebox(0,0)[lt]{\lineheight{1.25}\smash{\begin{tabular}[t]{l}$P$\end{tabular}}}}%
    \put(0.44304666,0.00923487){\color[rgb]{0,0,0}\makebox(0,0)[lt]{\lineheight{1.25}\smash{\begin{tabular}[t]{l}$\O$\end{tabular}}}}%
    \put(0.886916,0.00911821){\color[rgb]{0,0,0}\makebox(0,0)[lt]{\lineheight{1.25}\smash{\begin{tabular}[t]{l}$E$\end{tabular}}}}%
    \put(0,0){\includegraphics[width=\unitlength,page=1]{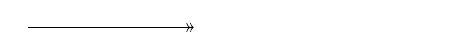}}%
    \put(0.19737706,0.04466546){\color[rgb]{0,0,0}\makebox(0,0)[lt]{\lineheight{1.25}\smash{\begin{tabular}[t]{l}$\obs$\end{tabular}}}}%
    \put(0.64474584,0.04201913){\color[rgb]{0,0,0}\makebox(0,0)[lt]{\lineheight{1.25}\smash{\begin{tabular}[t]{l}$\pred$\end{tabular}}}}%
    \put(0,0){\includegraphics[width=\unitlength,page=2]{Obspred.pdf}}%
  \end{picture}%
\endgroup%

\caption{\small
We assume that an experimental setup apt for a particular model of consciousness has been chosen for some class of physical systems $P$, wherein $p \in P$ represents the dynamics or configurations of a particular physical system. $\O$ then denotes all datasets that can arise from observations or measurements on $P$. Measuring the observables of $p$ maps to datasets $o \in \O$, which is denoted by the $obs$ correspondence. $E$ represents the mathematical description of experience given by the theory or model of consciousness under consideration. In the simplest case, this is just a set whose elements indicate whether a stimulus has been perceived consciously or not, but far more complicated structures can arise (e.g., in IIT). The correspondence $pred$ describes the process of prediction as a map from $\O$ to $E$.}
\label{fig:Experiments}
\end{figure}

Shown in Figure \ref{fig:Experiments} are the full set of terms needed to formally define how most contemporary theories of consciousness make predictions about experience. So far, what we have said is very general. Indeed, the force and generalizability of our argument comes from the fact that we do not have to define $pred$ explicitly for the various models we consider. It suffices that it exists, in some form or the other, for the models under consideration. 

It is crucial to note that predicting states of consciousness alone does not suffice to test a model of consciousness. Some have previously criticized theories of consciousness, IIT in particular, just based off of their counter-intuitive predictions. An example is the criticism that relatively simply grid-like networks have high $\Phi$ \citep{aaronson2014not, tononi2014scott}. However, debates about counter-intuitive predictions are not meaningful by themselves, since $\pred$ alone does not contain enough information to say whether a theory is true or false. The most a theory could be criticized for is either not fitting our own phenomenology or not being parsimonious enough, neither of which are necessarily violated by counter-intuitive predictions. For example, it may actually be parsimonious to assume that many physical systems have consciousness \citep{goff2017consciousness}. That is, speculation about acceptable predictions by theories of consciousness must implicitly rely on a comparative reference to be meaningful, and speculations that are not explicit about their reference are uninformative.

\subsection{Inferences}\label{sec:Inferences}

As discussed in the previous section, a theory is unfalsifiable given just predictions alone, and so $pred$ must be compared to something else. Ideally this would be the actual conscious experience of the system under investigation. However, as noted previously, the class of theories we focus on here assumes that experience itself is not part of the observables. For this reason, the experience of a system must be inferred separately from a theory's prediction to create a basis of comparison. Most commonly, such inferences are based on \emph{reports}. For instance, an inference might be based on an experimental participant reporting on the switching of some perceptually bistable image \citep{blake2014can} or on reports about seen vs. unseen images in masking paradigms \citep{alais2010visual}.

It has been pointed out that report in a trial may interfere with the actual isolation of consciousness, and there has recently been the introduction of so-called ``no-report paradigms'' \citep{tsuchiya2015no}. In these cases, report is first correlated to some autonomous phenomenon like optokinetic nystagmus (stereotyped eye movement), and then the experimenter can use this instead of the subject's direct reports to infer their experiences. Indeed, there can even be simpler cases where report is merely assumed: e.g., that in showing a red square a participant will experience a red square without necessarily asking the participant, since previously that participant has proved compos mentis. Similarly, in cases of non-humans incapable of verbal report, ``report'' can be broadly construed as behavior or output.

All these cases can be broadly described as being a case of inference off of some data. This data might be actual reports (like a participant's button pushes) or may be based off of physiological reactions (like no-report paradigms) or may be the outputs of a neural network or set of logic gates, such as the results of an image classification task \citep{lecun2015deep}. Therefore, the inference can be represented as a function, $\inf(o)$, between a dataset $o$ generated by observation or measurement of the physical system, and the set of postulated experiences in the model of consciousness, $E$:

\[
\inf: \O \rightarrow E \:.
\]

\begin{figure}
\centering
\begingroup%
  \makeatletter%
  \providecommand\color[2][]{%
    \errmessage{(Inkscape) Color is used for the text in Inkscape, but the package 'color.sty' is not loaded}%
    \renewcommand\color[2][]{}%
  }%
  \providecommand\transparent[1]{%
    \errmessage{(Inkscape) Transparency is used (non-zero) for the text in Inkscape, but the package 'transparent.sty' is not loaded}%
    \renewcommand\transparent[1]{}%
  }%
  \providecommand\rotatebox[2]{#2}%
  \newcommand*\fsize{\dimexpr\f@size pt\relax}%
  \newcommand*\lineheight[1]{\fontsize{\fsize}{#1\fsize}\selectfont}%
  \ifx\svgwidth\undefined%
    \setlength{\unitlength}{225.66016875bp}%
    \ifx\svgscale\undefined%
      \relax%
    \else%
      \setlength{\unitlength}{\unitlength * \real{\svgscale}}%
    \fi%
  \else%
    \setlength{\unitlength}{\svgwidth}%
  \fi%
  \global\let\svgwidth\undefined%
  \global\let\svgscale\undefined%
  \makeatother%
  \begin{picture}(1,0.29720736)%
    \lineheight{1}%
    \setlength\tabcolsep{0pt}%
    \put(-0.00514982,0.12986952){\color[rgb]{0,0,0}\makebox(0,0)[lt]{\lineheight{1.25}\smash{\begin{tabular}[t]{l}$P$\end{tabular}}}}%
    \put(0.44304666,0.12991772){\color[rgb]{0,0,0}\makebox(0,0)[lt]{\lineheight{1.25}\smash{\begin{tabular}[t]{l}$\O$\end{tabular}}}}%
    \put(0.886916,0.12980106){\color[rgb]{0,0,0}\makebox(0,0)[lt]{\lineheight{1.25}\smash{\begin{tabular}[t]{l}$E$\end{tabular}}}}%
    \put(0,0){\includegraphics[width=\unitlength,page=1]{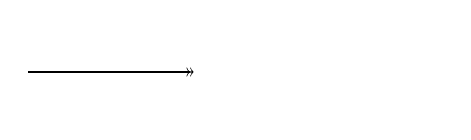}}%
    \put(0.19737706,0.16534831){\color[rgb]{0,0,0}\makebox(0,0)[lt]{\lineheight{1.25}\smash{\begin{tabular}[t]{l}$\obs$\end{tabular}}}}%
    \put(0,0){\includegraphics[width=\unitlength,page=2]{Obspredinf.pdf}}%
    \put(0.64847448,0.26017193){\color[rgb]{0,0,0}\makebox(0,0)[lt]{\lineheight{1.25}\smash{\begin{tabular}[t]{l}$\pred$\end{tabular}}}}%
    \put(0,0){\includegraphics[width=\unitlength,page=3]{Obspredinf.pdf}}%
    \put(0.65396183,0.00716431){\color[rgb]{0,0,0}\makebox(0,0)[lt]{\lineheight{1.25}\smash{\begin{tabular}[t]{l}$\inf$\end{tabular}}}}%
  \end{picture}%
\endgroup%

\caption{\small Two maps are necessary for a full experimental setup, one that describes a theory's predictions about experience ($pred$), another that describes the experimenter's inference about it ($inf$). Both map from a dataset $o \in O$ collected in an experimental trail to some subset of experiences described by the model, $E$.}
\label{fig:SetupAssumption}
\end{figure}

Defining $\inf$ as a function means that we assume that for every experimental dataset $o$, one single experience in $E$ is inferred during the experiment. Here we use a function instead of a correspondence for technical and formal ease, which does not affect our results: If two correspondences to the same space are given, one of them can be turned into a function.\footnote{If $\inf$ is a correspondence, one defines a new space $E'$ by $E' := \{ \inf(o) \, | \, o \in \O\}$. Every individual element of this space describes exactly what can be inferred from one dataset $o \in \O$, so that $\inf': \O \rightarrow E'$ is a function. The correspondence $\obs$
is then redefined, for every $e' \in E'$, by the requirement that $e' \in \obs'(o)$ iff $e \in \obs(o)$ for some $e \in e'$.
}
The $inf$ function is flexible enough to encompass both direct report, no-report, input/output analysis, and also assumed-report cases. It is a mapping that describes the process of inferring the conscious experience of a system from data recorded in the experiments. Both $inf$ and $pred$ are depicted in Figure~\ref{fig:SetupAssumption}.

It is worth noting that we have used here the same class $\O$ as in the definition of the prediction mapping $\pred$ above. This makes sense because the inference process also uses data obtained in experimental trials, such as reports by a subject. So both $pred$ and $inf$ can be described to operate on the same total dataset measured, even though they usually use different parts of this dataset (cf. below). 

\subsection{Falsification}\label{sec:Falsification}
We have now introduced all elements which are necessary to formally say what a falsification of a theory of consciousness is. To falsify a theory of consciousness requires mismatch between an experimenter's inference (generally based on report) and the predicted consciousness of the subject. In order to describe this, we consider some particular experimental trial, as well as $\inf$ and $\pred$.

\begin{Def}\label{def:falsi}
There is a \emph{falsification} at $o \in \O$ if we have
\begin{equation}\label{eq:Falsification}
\inf(o) \not \in \pred(o) \:.
\end{equation}
\end{Def}

This definition can be spelled out in terms of individual components of $E$. To this end, for any given dataset $o \in \O$, let $e_r := \inf(o)$ denote the experience that is being inferred, and let $e_p \in \obs(o)$ be one of the experiences that is predicted based off of some dataset. Then~\eqref{eq:Falsification} simply states that we have $e_p \neq e_r$ for all possible predictions $e_p$ $\in \obs(o)$. None of the predicted states of experience is equal to the inferred experience.

What does Equation~\eqref{eq:Falsification} mean? There are two cases which are possible. Either, the prediction based on the theory of consciousness is correct and the inferred experience is wrong. Or the prediction is wrong, so that in this case the model would be falsified. In short: Either the prediction process or the inference process is wrong. 

We remark that if there is a dataset $o$ on which the inference procedure $\inf$ or the prediction procedure $\pred$ cannot be used, then this dataset cannot be used in falsifying a model of consciousness. Thus, when it comes to falsifications, we can restrict to datasets $o$ for which both procedures are defined.

In order to understand in more detail what is going on if~\eqref{eq:Falsification} holds, we have to look into a single dataset $o \in \O$. This will be of use later.

\begin{figure}
\centering
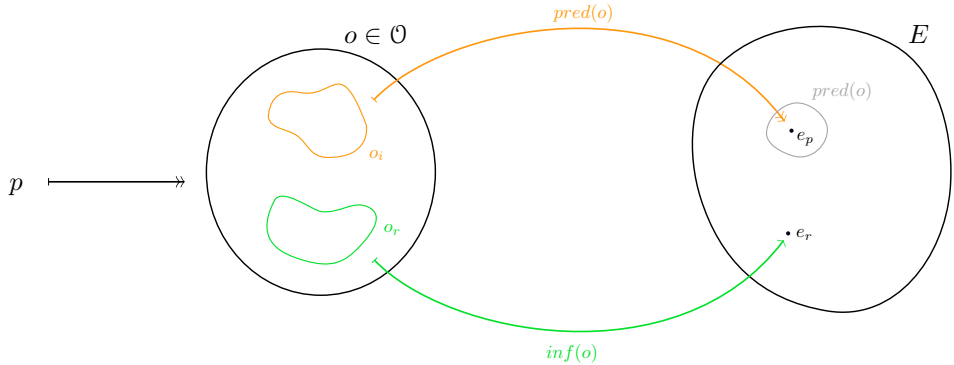
\caption{This figure represents the same setup as Figure~\ref{fig:SetupAssumption}.
The left circle depicts one single dataset $o$. $o_i$ (orange) is the part of the dataset used for prediction. $o_r$ (green) is the part of the dataset used for inferring the state of experience. Usually the green area comprises verbal reports or button presses, whereas the orange area comprises the data obtained from brain scans. The right circle depicts the experience space $E$ of a theory under consideration. $e_p$ denotes a predicted experience while $e_r$ denotes the inferred experience. Therefore, in total, to represent some specific experimental trial we use $p \in P$, $o \in \O$, $e_r \in E$ and $e_p \in E$, where $e_p \in \pred(o)$.
}
\label{fig:PeekElement}
\end{figure}

Generally, $\inf$ and $\obs$ will make use of different part of the data obtained in an experimental trial. E.g., in the context of IIT or GNW, data about the internal structure and state of the brain will be used for the prediction. This data can be obtained from an fMRI scan or EEG measurement. The state of consciousness on the other hand can be inferred from verbal reports. Pictorially, we may represent this as in Figure~\ref{fig:PeekElement}. We use the following notation:
\begin{itemize}[leftmargin=1cm]
\item[$o_i$] For a chosen dataset $o \in O$, we denote the part of the dataset which is used for the prediction process by $o_i$ (for `internal' data). This can be thought of as data about the internal workings of the system. We call $o_i$ the \emph{prediction data} in $o$.
\item[$o_r$] For a chosen dataset $o \in O$, we denote the part of the dataset which is used for inferring the state of experience by $o_r$ (for `report' data). We call it the \emph{inference data} in $o$.
\end{itemize}
Note that in both cases, the subscript can be read similarly as the notation for restricting a set. We remark that a different kind of prediction could be considered as well, where one makes use of the inverse of $\pred$. In Appendix~\ref{app:inverse}, we prove that this is in fact equivalent to the case considered here, so that Definition~\ref{def:falsi} indeed covers the most general situation.

\subsection{Summary}\label{sec:Summary}

In summary, for testing of a theory of consciousness we have introduced the following notion:
\begin{itemize}[leftmargin=1cm]
\item[$P$] denotes a class of physical systems that could have been tested, in principle, in the experiment under consideration, each in various different configurations. In most cases, every $p \in P$ thus describes a physical system in a particular state, dynamical trajectory, or configuration.
\item[$\obs$] is a correspondence which contains all details on how the measurements are set up and what is measured. It describes how measurement results (datasets) are determined by a system configuration under investigation. This correspondence is given, though usually not explicitly known, once a choice of measurement scheme has been made.
  
\item[$\O$] is the class of all possible datasets that can result from observations or measurements of the systems in the class $P$. Any single experimental trail results in a single dataset $o \in O$, whose data is used for making predictions based on the theory of consciousness and for inference purposes.
\item[$pred$] describes the process of making predictions by applying some theory of consciousness to a dataset $o$. It is therefore a mapping from $\O$ to $E$.
\item[$E$] denotes the space of possible experiences specified by the theory under consideration. 
The result of the prediction is a subset of this space, denoted as $\pred(o)$. Elements of this subset are denoted by $e_i$ and describe predicted experiences.
\item[$\inf$] describes the process of inferring a state of experience from some observed data, e.g. verbal reports, button presses or using no-report paradigms. Inferred experiences are denoted by $e_r$.

\end{itemize}

\section{\label{sec:substitution}The substitution argument}

Substitutions are changes of physical systems (i.e., the substitution of one for another) that leave the inference data invariant, but may change the result of the prediction process. A specific case of substitution, the unfolding of a reentrant neural network to a feed-forward one, was recently applied to IIT to argue that IIT cannot explain consciousness \citep{doerig2019unfolding}.

Here we show that, in general, the contemporary notion of falsification in the science of consciousness exhibits this fundamental flaw for almost all contemporary theories, rather than being a problem for a particular theory. This flaw is based on the independence between the data used for inferences about consciousness (like reports) and the data used to make predictions about consciousness. We discuss various responses to this flaw in Section~\ref{sec:objections}.

We begin by defining what a substitution is in Section~\ref{sec:defsubstitution}, show that it implies falsifications in Section~\ref{sec:subsfals}, and analyze the particularly problematic case of universal substitutions in Section~\ref{sec:universalsubst}. In Section~\ref{sec:exist}, we prove that universal substitutions exist if prediction and inference data are independent and give some examples of already-known cases.

\subsection{Substitutions}\label{sec:defsubstitution}

In order to define formally what a substitution is, we work with the inference content $o_r$ of a dataset $o$ as introduced in Section~\ref{sec:Falsification}. We first denote the class of all physical configurations which could have produced the inference content $o_r$ upon measurement by $P_{o_r}$. Using the correspondence $\obs$ which describes the relation between physical systems and measurement results, this can be defined as
\begin{equation}\label{eq:Defor}
P_{o_r} := \{ \, p \in P \, | \, o_r \in \obs(p) \, \} \: ,
\end{equation}
where $\obs(p)$ denotes all possible datasets that can be measured if the system $p$ is under investigation and where $o_r \in \obs(p)$ is a shorthand for $o \in \obs(p)$ with inference content $o_r$.

Any map of the form $S: P_{o_r} \rightarrow P_{o_r}$ takes a system configuration $p$ which can produce inference content~$o_r$ to another system's configuration $S(p)$ which can produce the same inference content. 
This allows us to define what a substitution is formally. In what follows, the $\circ$ indicates the composition of the correspondences $\obs$ and $\pred$ to give a correspondence from $P$ to $E$, which could also be denoted as $\pred(\obs(p))$,\footnote{
I.e., $\pred \circ \obs(p) = \{ e \in E \, | \ e \in \pred(o) \textrm{ for some } o \in \obs(p) \}$, it is the image under $\pred$ of the set $\obs(o)$.
}
and
$\cap$ denotes the intersection of sets.

\begin{Def}\label{def:orsubst}
There is a \emph{$o_r$-substitution} if there is a transformation $S: P_{o_r} \rightarrow P_{o_r}$ such that at least for one $p \in P_{o_r}$
\begin{equation}\label{eq:Subst}
\pred \circ \obs(p) \, \cap \, \pred \circ \obs(S(p)) = \emptyset  \:.
\end{equation}
\end{Def}

In words, a substitution requires there to be a transformation $S$ which keeps the inference data constant but changes the prediction of the system. So much in fact that the prediction of the original configuration $p$ and of the transformed configuration $S(p)$ are fully incompatible, i.e. there is no single experience $e$ which is contained in both predictions. Given some inference data $o_r$, an $o_r$-substitution then requires this to be the case for at least one system configuration $p$ that gives this inference data. In other words, the transformation $S$ is such that for at least one $p$, the \emph{predictions change completely}, while the inference content $o_r$ is preserved. 

\begin{figure}
\centering
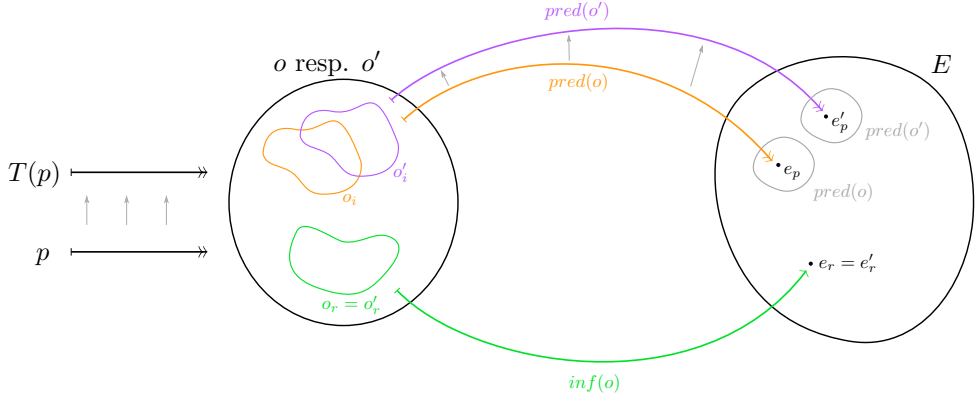
\caption{
This picture illustrates substitutions. Assume that some dataset $o$ with inference content $o_r$ is given. A substitution is a transformation $S$ of physical systems which leaves the inference content $o_r$ invariant but which changes the result of the prediction process. Thus whereas $p$ and $S(p)$ have the same inference content $o_r$, the prediction content of experimental datasets is different; different in fact to such an extend that the predictions of consciousness based on these datasets are incompatible (illustrated by the non-overlapping gray circles on the right). Here we have used that by definition of $P_{o_r}$, every $\tilde p \in P_{o_r}$ yields at least one dataset $o'$ with the same inference content as $o$ and have identified $o$ and $o'$ in the drawing.
}
\label{fig:Subst}
\end{figure}

A pictorial definition of substitutions is given in Figure~\ref{fig:Subst}. We remark that if $\pred$ and $\obs$ were functions, so that $\pred \circ \obs(p)$ only contained one element, Equation~\eqref{eq:Subst} would be equivalent to $\pred(\obs(p)) \neq \pred(\obs(S(p)))$.

We will find below that the really problematic case arises if there is an $o_r$-substitution for every possible inference content $o_r$. We refer to this case as a universal substitution.

\begin{Def}\label{def:univsubst}
There is a \emph{universal substitution} if there is an $o_r$-substitution $S_{o_r}: P_{o_r} \rightarrow P_{o_r}$ for every~$o_r$.
\end{Def}

We recall that according to the notation introduced in Section~\ref{sec:Falsification}, the inference content of any dataset $o \in \O$ is denoted by $o_r$ (adding the subscript $r$). Thus the requirement is that there is an $o_r$-substitution $S_{o_r}: P_{o_r} \rightarrow P_{o_r}$ for every inference data that can pertain in the experiment under consideration (for every inference data that is listed in~$\O$). The subscript $o_r$ of $S_{o_r}$ indicates that the transformation $S$ in Definition~\ref{def:orsubst} can be chosen differently for different $o_r$. Definition~\ref{def:univsubst} does not require there to be one single transformation that works for all $o_r$.

\subsection{Substitutions imply falsifications}\label{sec:subsfals}

The force of our argument comes from the fact that if there are substitutions, then this necessarily leads to mismatches between inferences and predictions. This is shown by the following lemma.
\begin{Lemma}\label{orsubstsomefalsi}
If there is a $o_r$-substitution, there is a falsification at some $o \in \O$.
\end{Lemma}

\Proof
Let $p$ be the physical system in Definition~\ref{def:orsubst} and define $p' = S(p)$. Let $o \in \obs(p)$ be a dataset of $p$ which has inference content $o_r$ and 
let $o'$ be a dataset of $p'$ which has the same inference content $o_r$, guaranteed to exist by the definition of $P_{o_r}$ in~\eqref{eq:Defor}.
Equation~\eqref{eq:Subst} implies that
\begin{equation}
\pred(o) \cap \pred(o') = \emptyset \:.
\end{equation}
Since, however, $o_r = o_r'$, we have $\inf(o) = \inf(o')$. Thus we have either
$\inf(o) \not \in \pred(o)$ or $\inf(o') \not \in \pred(o')$, or both. Thus there is either
a falsification at $o$, a falsification at $o'$, or both.
\QED

The last lemma shows that if there are substitutions, then there are necessarily falsifications. This might, however, not be considered too problematic, since it could always be the case that the model is right whereas the inferred experience is wrong. Inaccessible predictions are not unusual in science. A fully problematic case only pertains for universal substitutions, i.e., if there is an $o_r$-substitution for every inference content $o_r$ that can arise in an experiment under consideration.

\subsection{Universal substitutions imply complete falsification}\label{sec:universalsubst}

In Section~\ref{sec:Falsification}, we have defined falsifications for individual datasets $o \in \O$. 
Using the `insight view' of single datasets, we can refine this definition somewhat by relating it to the inference content only.

\begin{Def}\label{def:orfals}
There is an $o_r$-falsification if there is a falsification for some $o \in \O$ which has inference content~$o_r$.
\end{Def}

This definition is weaker than the original definition, because among all datasets which have inference content $o_r$, only one needs to exhibit a falsification. Using this notion, the next lemma specifies the exact relation between substitutions and falsifications.

\begin{Lemma}\label{lem:orsuborfals}
If there is an $o_r$-substitution, there is an $o_r$-falsification.
\end{Lemma}

\Proof
This lemma follows directly from the proof of Lemma~\ref{orsubstsomefalsi} because the datasets $o$ and $o'$ used in that proof both have inference content $o_r$.
\QED

This finally allows us to show our first main result. It shows that if a universal substitution exists, the theory of consciousness under consideration is falsified. We explain the meaning of this proposition after the proof.

\begin{Prp}\label{prp:universalsubs}
If there is a universal substitution, there is an $o_r$-falsification \emph{for all possible} inference contents $o_r$.
\end{Prp}

\Proof
By definition of universal substitution, there is an $o_r$-substitution for every $o_r$. Thus the claim follows directly from Lemma~\ref{lem:orsuborfals}.
\QED

In combination with Definition~\ref{def:orfals}, this proposition states that for every possible report (or any other type of inference procedure, cf. our use of terminology in Section~\ref{sec:Falsification}), there is a dataset $o$ which contains the report's data and for which we have 
\begin{equation}\label{eq:illustrateprp}
\inf(o_r) \notin \pred(o) \:,
\end{equation}
where we have slightly abused notation in writing $\inf(o_r)$ instead of $\inf(o)$ for clarity.
This implies that one of two cases needs to pertain:
Either at least one of the inferred experiences $\inf(o_r)$ is correct, in which case the corresponding prediction is wrong and the theory needs to be considered falsified.
The only other option is that for \emph{all} inference contents $o_r$, the prediction $\pred(o)$ is correct, which qua~\eqref{eq:illustrateprp} implies that no single inference $\inf(o_r)$ points at the correct experience, so that the inference procedure is completely wrong. This shows that Proposition~\ref{prp:universalsubs} can equivalently be stated as follows.

\begin{Prp}\label{prp:theoryfalse}
If there is a universal substitution, either every single inference operation is wrong or the theory under consideration is already falsified.
\end{Prp}

Next, we discuss under which circumstances a universal substitution exists.

\subsection{When does a universal substitution exist?}\label{sec:exist}

In the last section, we have seen that if a universal substitution exists, this has strong consequences. In this section, we discuss under what conditions universal substitutions exist.

\subsubsection{Theories need to be minimally informative}\label{sec:minimallyinformative}
We have taken great care above to make sure that our notion of prediction is compatible with incomplete or noisy datasets. This is the reason why $\pred$ is a correspondence, the most general object one could consider. For the purpose of this section, we add a gentle assumption which restricts $\pred$ slightly: we assume that every prediction carries at least a minimal amount of information. In our case, this means that for every prediction $\pred(o)$, there is at least one other prediction $\pred(o')$ which is different from $\pred(o)$. Put in simple terms, this means that we don't consider theories of consciousness which have only a single prediction.

In order to take this into account, for every $o \in \O$, we define $\bar o := \obs( \obs^{-1} (o) )$, which comprises exactly all those datasets which can be generated by physical systems $p$ that also generate $o$. 
When applying our previous definitions, this can be fleshed out as
\begin{equation}\label{eq:closureo}
\bar o = \{\, o' \, | \, \exists \, p \textrm{ such that }  o \in \obs(p) \textrm{ and } o' \in \obs(p) \, \} \:.
\end{equation}
Using this, we can state our \emph{minimal information assumption} in a way that is compatible with the general setup displayed in Figure~\ref{fig:SetupAssumption}:

We assume that the theories of consciousness under consideration are \emph{minimally informative} in that for every $o \in \O$, there exists an $o' \in \O$ such that
\begin{equation}\label{eq:minimalinformation}
    \pred(\bar o) \cap \pred(\bar o') = \emptyset \:.
\end{equation}

\subsubsection{Inference and prediction data are independent}

We have already noted, that in most experiments, the prediction content $o_i$ and inference content $o_r$ consist of different parts of a dataset. What is more, they are usually assumed to be independent, in the sense that changes in $o_i$ are possible while keeping $o_r$ constant. This is captured by the next definition.

\begin{Def}\label{def:independence}
\emph{Inference and prediction data are independent} if for any $o_i$, $o_i'$ and $o_r$, there is a variation
\begin{equation}\label{eq:def-indep}
    \nu: P \rightarrow P
\end{equation}
such that $o_i \in \obs(p)$, $o_i' \in \obs(\nu(p))$ but $o_r \in \obs(p)$ and $o_r \in \obs(\nu(p))$
for some $p \in P$.
\end{Def}

Here, we use the same shorthand as in~\eqref{eq:Defor}. For example, the requirement $o_i \in \obs(p)$ is a shorthand for there being an $o \in \obs(p)$ which has prediction content $o_i$. The variation $\nu$ in this definition is a variation in~$P$, which describes physical systems which could, in principle, have been realized in an experiment (cf. Section~\ref{sec:Summary}). We note that a weaker version of this definition can be given which still implies our results below, cf. Appendix~\ref{app:weak}. Note that if inference and prediction data are not independent, e.g. because they have a common cause, problems of tautologies loom large, cf. Section~\ref{sec:objections}. Throughout the text we often refer to Definition~\ref{def:independence} simply as ``independence''.

\subsubsection{Universal substitutions exist}

Combining the last two sections, we can now prove that universal substitutions exist.

\begin{Prp}\label{prp:univexists}
If inference and prediction data are independent, universal substitutions exist.
\end{Prp}

\Proof
To show that a universal substitution exists, we need to show that for every $o \in \O$, an $o_r$-substitution exists (Definition~\ref{def:orsubst}). Thus assume that an arbitrary $o \in \O$ is given. The minimal information assumption guarantees that there is an $o'$ such that Equation~\eqref{eq:minimalinformation} holds. As before, we denote the prediction content of $o$ and $o'$ by $o_i$ and $o_i'$, respectively, and the inference content of $o$ by $o_r$.

Since inference and prediction data are independent, there exists a $p \in P$ as well as a $\nu: P \rightarrow P$ such that $o_i \in \obs(p)$, $o_i' \in \obs(\nu(p))$, $o_r \in \obs(p)$ and $o_r \in \obs(\nu(p))$. By Definition~\eqref{eq:closureo}, the first two of these four conditions imply that $\obs(p) \subset \bar o$ and $\obs(\nu(p)) \subset \bar o'$. Thus Equation~\eqref{eq:minimalinformation} applies and allows us to conclude that 
\[
\pred(\obs(p)) \cap \pred(\obs(\nu(p)) = \emptyset \:.
\]

Via Equation~\eqref{eq:Defor}, the latter two of the four conditions imply that $p \in P_{o_r}$ and $\nu(p) \in P_{o_r}$. Thus we may restrict $\nu$ to $P_{o_r}$ to obtain a map
\[
S: P_{o_r} \rightarrow P_{o_r} \:,
\]
which in light of the first part of this proof exhibits at least one $p \in P_{o_r}$ which satisfies~\eqref{eq:Subst}. Thus we have shown that an $o_r$-substitution exists. Since $o$ was arbitrary, it follows that a universal substitution exists.
\QED

The intuition behind this proof is very simple.
In virtue of our assumption that theories of consciousness need to be minimally informative, for any dataset $o$, there is another dataset $o'$ which makes a non-overlapping prediction. But in virtue of inference and prediction data being independent, we can find a variation that changes the prediction content as prescribed by $o$ and $o'$, but keeps the inference content constant. This suffices to show that there exists a transformation $S$ as required by the definition of a substitution.

Combining this result with Proposition~\ref{prp:theoryfalse}, we finally can state our main theorem.
\begin{Thm}\label{mainthm}
If inference and prediction data are independent, either every single inference operation is wrong or the theory under consideration is already falsified.
\end{Thm}

\Proof
The theorem follows by combining Proposition~\ref{prp:univexists} and Proposition~\ref{prp:theoryfalse}.
\QED

In the next section, we give several examples of universal substitutions, before discussing various possible responses to our result in Section~\ref{sec:objections}.

\subsubsection{Examples of data independence}\label{sec:examples-independence}

Our main theorem shows that testing a theory of consciousness will necessarily lead to its falsification if inference and prediction data are independent (Definition~\ref{def:independence}), and if at least one single inference can be trusted (Theorem~\ref{mainthm}). In this section, we give several examples that illustrate the independence of inference and prediction data. We take report to mean output, behavior, or verbal report itself and assume that prediction data derives from internal measurements.

\textit{Artificial neural networks}. ANNs, particularly those trained using deep learning, have grown increasingly powerful and capable of human-like performance \citep{lecun2015deep, bojarski2016end}. For any ANN, report (output) is a function of node states. Crucially, this function is non-injective, i.e. some nodes are not part of the output. E.g., in deep learning, the report is typically taken to consist of the last layer of the ANN, while the hidden layers are not taken to be part of the output.
Correspondingly, for any given inference data, one can construct a ANN with arbitrary prediction data by adding nodes, changing connections and changing those nodes which are not part of the output.  Put differently, one can always substitute a given ANN with another with different internal observables but identical or near-identical reports. From a mathematical perspective it is well-known that both feed-forward ANNs and recurrent ANNs can approximate any given function \citep{ hornik1989multilayer, schafer2006recurrent}. Since reports are just some function, it follows that there are viable universal substitutions.

A special case thereof is the unfolding transformation considered in~\cite{doerig2019unfolding} in the context of IIT. The arguments in this paper constitute a proof of the fact that for ANNs, inference and prediction data are independent (Definition~\ref{def:independence}). Crucially, our main theorem shows that this has implications for all minimally informative theories of consciousness. A similar result (using a different characterization of theories of consciousness than minimally informative) has been shown in~\citep{kleiner2019empirical}.

\textit{Universal computers}. Turing machines are extremely different in architecture than ANNs. Since they are capable of universal computation \citep{turing1937computable} they should provide an ideal candidate for a universal substitution. Indeed, this is exactly the reasoning behind the Turing test of conversational artificial intelligence \citep{turing2009computing}. Therefore, if one believes it is possible for a sufficiently fast Turing machine to pass the Turing test, one needs to accept that substitutions exist. Notably, Turing machines are just one example of universal computation, and there are other instances of different parameter spaces or physical systems that are capable thereof, such as cellular automata \citep{wolfram1984cellular}.

\textit{Universal intelligences}. There are models of universal intelligence that allow for maximally intelligent behavior across any set of tasks \citep{hutter2003gentle}. For instance, consider the AIXI model, the gold-standard for universal intelligence, which operates via Solomonoff induction \citep{solomonoff1964formal, hutter2004universal}. The AIXI model generates an optimal decision making over some class of problems, and methods linked to it have already been applied to a range of behaviors, such as creating ``AI physicists'' \citep{wu2019toward}. Its universality indicates it is a prime candidate for universal substitutions. Notably, unlike a Turing machine, it avoids issues of precisely how it is accomplishing universal substitution of report, since the algorithm that governs the AIXI model behavior is well-described and ``relatively'' simple.

The above are all real and viable classes of systems that are used everyday in science and engineering which all provide different viable universal substitutions if inferences are based on reports or outputs. They show that in normal experimental setups such as the ones commonly used in neuroscientific research into consciousness \citep{frith1999neural}, inference and prediction data are indeed independent, and dependency is not investigated nor properly considered. It is always possible to substitute the physical system under consideration with another that has different internal observables, and therefore different predictions, but similar or identical reports. Indeed, recent research in using the work introduced in this work shows that even different spatiotemporal models of a system can be substituted for one another, leading to falsification \citep{hanson2020formalizing}. We have not considered possible but less reasonable examples of universal substitutions, like astronomically-large look-up ledgers of reports.

As an example of our Main Theorem~\ref{mainthm}, we consider the case of IIT. Since the theory is normally applied in Boolean networks, logic gates, or artificial neural networks, one usually takes report to mean ``output.'' In this case, it has already been proven that systems with different internal structures and hence different predicted experiences, can have identical input/output (and therefore identical reports or inferences about report) \citep{albantakis2019causal}. To take another case: within IIT it has already been acknowledged that a Turing machine may have a wildly different predicted contents of consciousness for the same behavior or reports \citep{koch2019feeling}. Therefore, data independence during testing has already been shown to apply to IIT under its normal assumptions. 

\section{\label{sec:dependence}Inference and prediction data are strictly dependent}

An immediate response to our main result showing that many theories suffer from \textit{a priori} falsification would be to claim that it offers support of theories which define conscious experience in terms of what is accessible to report. This is the case, e.g., for behaviourist theories of consciousness but might arguably also be the case for some interpretations of global workspace theory or fame in the brain proposals. In this section, we show that this response is not valid, as theories of this kind, where inference and prediction data are strictly dependent, are unfalsifiable.

In order to analyse this case, we first need to specifically outline how theories can be pathologically unfalsifiable. Clearly, the goal of the scientific study as a whole is to find, eventually, a theory of consciousness that are empirically adequate and therefore corroborated by \textit{all} experimental evidence. Therefore, not being falsified in experiments is a necessary condition (though not sufficient) any purportedly ``true'' theory of consciousness needs to satisfy. Therefore, not being falsifiable by the set of possible experiments per se is not a bad thing. We seek to distinguish this from cases of unfasifiability due to pathological assumptions that underlie a theory of consciousness, assumptions which render an experimental investigation meaningless. Specifically, a pathological dependence between inferences and predictions leads to theories which are unfalsifiable.

Such unfalsifiable theories can be identified neatly in our formalism. To see how, recall that~$\O$ denotes the class of all datasets that can result from an experiment investigating the physical systems in the class~$P$. Put differently, it contains all datasets that could, in principle, appear when probed in the experiment. This is \emph{not} the class of all possible datasets of type $\O$ one can think of. Many datasets which are of the same form as elements of $\O$ might simply not arise in the experiment under consideration. We denote the class of all possible datasets as:
\[
\Obar: \textrm{ All possible datasets of type }\O \: .
\]
Intuitively, in terms of possible worlds semantics, $\O$ describes the datasets which could appear, for the type of experiment under consideration, in the actual world. $\Obar$, in contrast, describes the datasets which could appear in this type of experiment in any possible world. For example, $\Obar$ contains datasets which can only occur if consciousness attaches to the physical in a different way than it actually does in the actual word.

By construction, $\O$ is a subset of $\Obar$, which describes which among the possible datasets actually arises across experimental trials. Hence, $\O$ also determines which theory of consciousness is compatible with (i.e. not falsified by) experimental investigation. However, $\Obar$ defines all possible data sets independent of any constraint by real empirical results, that is, all possible imaginable data sets.

Introduction of $\Obar$ allows us to distinguish the pathological cases of unfalsifiability mentioned above. Whereas any purportedly true theory should only fail to be falsified with respect to the experimental data $\O$, a pathological unfalsifiability pertains if a theory cannot be falsified at all, i.e. over $\Obar$. This is captured by the following definition.

\begin{Def}\label{def:infpreddep}
A theory of consciousness which does not have a falsification over $\Obar$ is \emph{empirically unfalsifiable}.
\end{Def}

\noindent Here, we use the term `empirically unfalsifiable' to highlight and refer to the pathological notion of unfalsifiability. Intuitively speaking, a theory which satisfies this definition appears to be true independently of any experimental investigation, and without the need for any such investigation. Using $\Obar$, we can also define the notion of strict dependence in a useful way.

\begin{Def}
Inference and prediction data are \emph{strictly dependent} if there is a function $f$ such that for any $o \in \Obar$,
we have $o_i = f(o_r)$.
\end{Def}

\newcommand{\Oall}{\O_{\textrm{all}}}
\newcommand{\Oexp}{\O_{\textrm{exp}}}

This definition says that there exists a function $f$ which for every possible inference data $o_r$ allows to deduce the prediction data $o_i$. We remark that the definition refers to $\Obar$ and not $\O$, as the dependence of inference and prediction considered here holds by assumption and is not simply asserting a contingency in nature.

The definition is satisfied, for example, if inference data is equal to prediction data, i.e. if $o_i = o_r$, where~$f$ is simply the identity. This is the case, e.g., for behaviourist theories \citep{skinner1938behavior} of consciousness, where consciousness is equated directly with report or behavior, or for precursors of functionalist theories of consciousness that are based on behavior or input/output \citep{putnam1960minds}. The definition is also satisfied in the case where prediction data is always a subset of the inference data: 
\begin{equation}\label{eq:oisubeqor}
o_i \subseteq o_r \:.
\end{equation}
Here, $f$ is simply the restriction function. This arguably applies to global workspace theory \citep{baars2005global}, the ``attention schema'' theory of consciousness \citep{graziano2015attention} or ``fame in the brain'' \citep{dennett1993consciousness} proposals. 

In all these cases, consciousness is generated by -- and hence needs to be predicted via -- what is accessible to report or output. In terms of Block's distinction between phenomenal consciousness and access consciousness~\citep{block1996can}, Equation~\eqref{eq:oisubeqor} holds true whenever a theory of consciousness is under investigation 
where access consciousness determines phenomenal consciousness.

Our second main theorem is the following.

\begin{Thm}\label{mainthm2}
If a theory of consciousness implies that inference and prediction data are strictly dependent, then it is either already falsified or empirically unfalsifiable.
\end{Thm}

\Proof
To prove the theorem, it is useful to consider the inference and prediction content of datasets explicitly. The possible pairings that can occur in an experiment are given by
\begin{equation}\label{eq:Oexp}
\Oexp := \{ \, (o_i,o_r) \ | \ o \in \O \, \} \:,
\end{equation}
where we have again used our notation that $o_i$ denotes the prediction data of $o$, and similar for $o_r$.
To define the possible pairings that can occur in $\Obar$, we let $\O_i$ denote the class of all prediction contents that arise in~$\O$, and $\O_r$ denote the class of all inference contents that arise in $\O$. The set of all conceivable pairings is then given by
\begin{align}\label{eq:Oall}
\Oall :=&  \{ \, (o_i,o_r') \ | \ o \in \O, \ o' \in \O\} \\
=& \{ \, (o_i,o_r') \ | \ o_i \in \O_i, \ o_r' \in \O_r \, \} \:.
\end{align}
Crucially, here, $o_i$ and $o_r'$ do not have to be part of the same dataset $o$.
Combined with Definition~\ref{def:falsi}, we conclude that there is a falsification over $\Obar$ if for some $(o_i,o_r') \in \Oall$, we have $\inf(o) \notin \pred(o')$, and there is a falsification over $\O$ if for some
$(o_i,o_r) \in \Oexp$, we have $\inf(o) \notin \pred(o)$.

Next we show that if inference and prediction data are strictly dependent, then $\Oall=\Oexp$ holds.
We start with the set $\Oall$ as defined in~\eqref{eq:Oall}. 
Expanding this definition in words, it reads
\begin{equation}\label{eq:proofunfasi}
\Oall = \{ \, (d_i,d_r) \ | \ \exists \, o \in \O \textrm{ such that } d_r=o_r \textrm{ and } \exists \, \tilde o \in \O \textrm{ such that } d_i=\tilde o_i \, \} \:,
\end{equation}
where we have symbols $d_i$ and $d_r$ to denote prediction and inference data independently of any dataset $o$.

Assume that the first condition in this expression, $d_r = o_r$ holds for some $o \in \O$. Since inference and prediction data are strictly dependent, we have $d_i = f(d_r)$. Furthermore, for the same reason, the prediction content $o_i$ of the dataset $o$ satisfies $o_i = f(o_r)$. Applying the function $f$ to both sides of the first condition gives $f(d_r) = f(o_r)$, which thus in turn implies $o_i = d_i$. This means that the $o$ that satisfies the first condition
in~\eqref{eq:proofunfasi} automatically also satisfies the second condition. Therefore, due to inference and prediction data being strictly dependent,~\eqref{eq:proofunfasi} is equivalent to
\begin{equation}\label{eq:proofunfasi2}
\Oall = \{ \, (d_i,d_r) \ | \ \exists \, o \in \O \textrm{ such that } d_r=o_r \textrm{ and } d_i= o_i \, \} \:.
\end{equation}
This, however, is exactly $\Oexp$ as defined in~\eqref{eq:Oexp}. Thus we conclude that if inference and prediction data are strictly dependent, $\Oall=\Oexp$ necessarily holds.

Returning to the characterisation of falsification in terms of $\Oexp$ and $\Oall$ above, what we have just found implies that there is a falsification over $\O$ if and only if there is a falsification over $\Obar$. Thus either there is a falsification over $\O$, in which case the theory is already falsified or there is no falsification over $\Obar$, in which case the theory under consideration is empirically unfalsifiable.
\QED

The gist of this proof is that if inference and prediction data are strictly dependent, then as far as the inference and prediction contents go, $\O$ and $\Obar$ are the same. I.e, the experiment does not add \textit{anything} to the evaluation of the theory. It is sufficient to know only all possible datasets to decide whether there is a falsification. In practise, this would mean that knowledge of the experimental design (which reports are to be collected, on the one hand, which possible data a measurement device can produce, one the other) is sufficient to evaluate the theory, which is clearly at odds with the role of empirical evidence required in any scientific investigation. Thus such theories are empirically unfalsifiable.

To give an intuitive example of the theorem, let us examine a theory that uses the information accessible to report in a system to predict conscious experience (perhaps this information is ``famous'' in the brain or is within some accessible global workspace). In terms of our notation, we can assume that $o_r$ denotes everything that is accessible to report, and $o_i$ denotes that part which is used by the theory to predict conscious experience. Thus in this case we have $o_i \subseteq o_r$. Since the predicted contents are always part of what can be reported, there can never be any mismatch between reports and predictions. However, this is not only the case for $\Oexp$ but also, in virtue of the theory's definition, for all possible datasets, i.e., $\Oall$. Therefore such theories are empirically unfalsifiable. Experiments add no information to whether the theory is true or not, and such theories are empirically uninformative or tautological.

\section{\label{sec:objections}Objections}

In this section, we discuss a number of possible objections to our results.

\subsection{Restricting inferences to humans only}

The examples given in Section~\ref{sec:examples-independence} show that data independence holds during the usual testing setups. This is because prima facie it seems reasonable to base inferences either on report capability or intelligent behavior in a manner agnostic of the actual physical makeup of the system. Yet this entails independence, so in these cases our conclusions apply.

One response to our results might be to restrict all testing of theories of consciousness solely to humans. In our formalisms this is equivalent to making the strength of inferences based not on reports themselves but on an underlying biological homology. Such an $inf$ function may still pick out specific experiences via reports, but the weight of the inference is carried by homology rather than report or behavior. This would mean that the substitution argument does not significantly affect consciousness research, as reports of non-human systems would simply not count. Theories of consciousness, so this idea goes, would be supported by abductive reasoning from testing in humans alone.

Overall there are strong reasons to reject this restriction of inferences. One significant issue is that this objection is equivalent to saying that reports or behavior in non-humans carry \textit{no} information about consciousness, an incredibly strong claim. If non-humans contradicted a theory (like a complex organism acting in pain while a theory predicted a lack of pain) the theory would be presumed to be correct above any behavior or report, meaning that abductive application of the theory ignores the fact that this sort of abductive reasoning should actually falsify the theory. Indeed, this is highly problematic for consciousness research which often uses non-human animal models \citep{boly2013consciousness}. For instance, cephalopods are among the most intelligent animals yet are quite distant on the tree of life from humans and have a distinct neuroanatomy, and still are used for consciousness research \citep{mather2008cephalopod}. Even in artificial intelligence research, there is increasing evidence that deep neural networks produced brain-like structures such as grid cells, shape tuning, and visual illusions, and many others \citep{richards2019deep}. Given these similarities, it becomes questionable why the strength of inferences should be based on homology instead of capability of report or intelligence. 

What is more, restricting inferences to humans alone is unlikely to be sufficient to avoid our results. Depending on the theory under consideration, data independence might exist just in human brains alone. That is, it is probable that there are transformations (as in Equation \eqref{eq:def-indep}) available within the brain wherein $o_r$ is fixed but $o_i$ varies. This is particularly true once one allows for interventions on the human brain by experimenters, such as perturbations like transcranial magnetic stimulation, which is already used in consciousness research \citep{rounis2010theta, napolitani2014transcranial}. 

For these reasons this objection does not appear viable. At minimum it is clear that if the objection were taken seriously, it would imply significant changes to consciousness research which would make the field extremely restricted with strong \textit{a priori} assumptions.

\subsection{Reductio ad absurdum}\label{sec:reductio}

Another hypothetical objection to our results is to argue that they could just as well be applied to scientific theories in other fields. If this turned out to be true this wouldn't imply our argument is necessarily incorrect. But the fact that other scientific theories do not seem especially problematic with regard to falsification would generate the question of whether some assumption is illegitimately strong. In order to address this, we explain which of our assumptions is specific to theories of consciousness and wouldn't hold when applied to other scientific theories. Subsequently, we give an example to illustrate this point.

The assumption in question is that $\O$, the class of all datasets that can result from observations or measurements of a system, is determined by the physical configurations in $P$ alone. I.e., every single dataset~$o$, including both its prediction content $o_i$ and its inference content $o_r$, is determined by $p$, and \emph{not} by a conscious experience in $E$. In Figure~\ref{fig:SetupAssumption}, this is reflected in the fact that there is an arrow from $P$ to $\O$, but no arrow from $E$ to $\O$.

This assumption expresses the standard paradigm of testing theories of consciousness in neuroscience, according to which both the data used to predict a state of consciousness and the reports of a system are determined by its physical configuration alone. This, in turn, may be traced back to consciousness' assumed subjective and private nature, which implies that any empirical access to states of consciousness in scientific investigations is necessarily mediated by a subject's reports, and to general physicalist assumptions.

This is different from experiments in other natural sciences. If there are two quantities of interest whose relation is to be modelled by a scientific theory, then in all reasonable cases there are two \textit{independent} means of collecting information relevant to a test of the theory, one providing a dataset that is determined by the first quantity, and one providing a dataset that is determined by the second quantity.

Consider, as an example, the case of temperature $T$ and its relation to microphysical states. To apply our argument, the temperature $T$ would replace the experience space $E$ and $p$ would denote a microphyiscal configuration. In order to test any particular theory about how temperature is determined by microphysical states, one would make use of two different measurements. The first measurement would access the microphysical states and would allow measurement of, say, the mean kinetic energy (if that's what the theory under consideration utilizes). This first measurement would provide a dataset $o_m$ that replaces the prediction data $o_i$ above. For the second measurement, one would use a thermometer or some other measuring device to obtain a dataset $o_t$ that replaces our inference data $o_r$ above. Comparison of the inferred temperature with the temperature that is predicted based on $o_m$ would allow testing of the theory under consideration. These independent means provide independent access to each of the two datasets in question. Combining $o_m$ and $o_t$ in one dataset $o$, the diagrammatic representation is
\[  P \longrightarrow \O \longleftarrow T \:, \]
which differs from the case of theories of consciousness considered here, wherein the physical system determines both datasets.

\subsection{\label{sec:theories_all_unfalsifiable}Theories could be based on phenomenology}

Another response to the issue of independence/dependence identified here is to propose that a theory of consciousness may not have to be falsified but can be judged by other characteristics. This is reminiscent of ideas put forward in connection with String Theory, which some have argued can be judged by elegance or parsimony alone \citep{carroll2018beyond}.

In addition to elegance and parsimony, in consciousness science, one could in particular consider a theory's fit with phenomenology, i.e. how well a theory describes the general structure of conscious experience. Examples of theories that are constructed based on a fit with phenomenology are recent versions of IIT \citep{oizumi2014phenomenology} or any view that proposes developing theories based on isomorphisms between the structure of experiences and the structure of physical systems or processes \citep{tsuchiya2019reply}.

It might be suggested that phenomenological theories might be immune to aspects of the issues we outline in our results \citep{negro2020phenomenology}. We emphasize that in order to avoid our results, and indeed the need for any experimental testing at all, a theory constructed from phenomenology has to be \emph{uniquely derivable} from conscious experience. However, to date, no such derivation exists, as phenomenology seems to generally underdetermine the postulates of IIT \citep{bayne2018axiomatic, barrett2019phi}, and because it is unknown what the scope and nature of non-human experience is. Therefore theories based on phenomenology can only confidently identify systems with human-like conscious experiences and cannot currently do so uniquely. Thus they cannot avoid the need for testing. 

As long as no unique and correct derivation exists across the space of possible conscious experiences, the use of experimental tests to assess theories of consciousness, and hence our results, cannot be avoided.

\subsection{Rejecting falsifiability}

Another response to our findings might be to deny the importance of falsifications within the scientific methodology. Such responses may reference a Lakatosian conception of science, according to which science does not proceed by discarding theories immediately upon falsification, but instead consists of \emph{research programs} built around a family of theories \citep{lakatos1980methodology}. These research programs have a \emph{protective belt} which consists of non-essential assumptions that are required to make predictions, and which can easily be modified in response to falsifications, as well as a \emph{hard core} that is immune to falsifications. Within the Lakatosian conception of science research programs are either progressive or degenerating based on whether they can ``anticipate theoretically novel facts in its growth'' or not \citep{lakatos1980methodology}. 

It is important to note, however, that Lakatos does not actually break with falsificationism. This is why Lakatos description of science is often called ``refined falsificationism'' in philosophy of science \citep{radnitzky1991refined}. Thus cases of testing theories' predictions remain relevant in a Lakatosian view in order to distinguish between progressive and degenerating research programs. Therefore our results generally translate into this view of scientific progress. In particular, Theorem~\ref{mainthm} shows that for every single inference procedure that is taken to be valid, there exists a system for which the theory makes a wrong prediction. This implies necessarily that a research program is degenerating. That is, independence implies that there is always an available substitution that can falsify any particular prediction coming from the research program.

\section{Conclusion}\label{sec:conclusion}

In this paper, we have subjected the usual scheme for testing theories of consciousness to a thorough formal analysis. We have shown that there appear to be deep problems inherent in this scheme which need to be addressed.

Crucially, in contrast to other similar results \citep{doerig2019unfolding}, we do not put the blame on individual theories of consciousness, but rather show that a key assumption that is usually being made is responsible for the problems: an experimenter's inference about consciousness and a theory's predictions are generally implicitly assumed to be independent during testing across contemporary theories. As we formally prove, if this independence holds, substitutions or changes to physical systems are possible that falsify any given contemporary theory. Whenever there is an experimental test of a theory of consciousness on some physical system which does not lead to a falsification, there necessary exists another physical system which, if it had been tested, would have produced a falsification of that theory. We emphasize that this problem does not only affect one particular type of theory, for example those based on causal interactions like IIT; theorems apply to all contemporary neuroscientific theories of consciousness if independence holds.

In the second part of our results, we examine the case where independence doesn't hold. We show that if an experimenter's inferences about consciousness and a theory's predictions are instead considered to be strictly dependent, empirical unfalsifiability follows, which renders any type of experiment to test a theory uninformative. This affects all theories wherein consciousness is predicted off of reports or behavior (such as behaviorism), theories based off of input/output functions, and also theories that equate consciousness with on accessible or reportable information.

Thus theories of consciousness seem caught between between Scylla and Charybdis, requiring delicate navigation. In our opinion there may only be two possible paths forward to avoid these dilemmas, which we briefly outline below. Each requires a revision of the current scheme of testing or developing theories of consciousness.

\paragraph{Lenient dependency.} When combined, our main theorems show that both independence and strict dependence of inference and prediction data are problematic and thus neither can be assumed in an experimental investigation. This raises the question of whether there are reasonable cases where inference and prediction are dependent, but not strictly dependent.

\textit{A priori}, in the space of possible relationships between inference and prediction data, there seems to be room for relationships that are neither independent (Section~\ref{sec:substitution}) nor strictly dependent (Section~\ref{sec:dependence}). We define this relationships of this kind as cases of \textit{lenient dependency}. No current theory or testing paradigm that we know of satisfies this definition. Yet cases of lenient dependency cannot be excluded to exist. Such cases would technically not be beholden to either Theorem~\ref{mainthm} or Theorem~\ref{mainthm2}.

There seems to be two general possibilities of how lenient dependencies could be built. On the one hand, one could hope to find novel forms of inference that allow to surpass the problems we have identified here. This would likely constitute a major change in the methodologies of experimental testing of theories of consciousness. On the other hand, another possibility to attain lenient dependence would be to construct theories of consciousness which yield prediction functions that are designed to explicitly have a leniently dependent link to inference functions. This would likely constitute a major change in constructing theories of consciousness.

\paragraph{Physics is not causally closed.} Another way to avoid our conclusion is to only consider theories of consciousness which do not describe the physical as causally closed \citep{kim1998mind}.  That is, the presence or absence of a particular experience itself would have to make a difference to the configuration, dynamics, or states of physical systems above and beyond what would be predicted with just information about the physical system itself. If a theory of consciousness does not describe the physical as closed, a whole other range of predictions are possible: predictions which concern the physical domain itself, e.g., changes in the dynamics of the system which depend on the dynamics of conscious experience. These predictions are not considered in our setup and may serve to test a theory of consciousness without the problems we have explored here.

\section{\label{sec:acknowledgments}Acknowledgments}
We would like to thank David Chalmers, Ned Block, and the participants of the NYU philosophy of mind discussion group for valuable comments and discussion. Thanks also to Ryota Kanai, Jake Hanson, Stephan Sellmaier, Timo Freiesleben, Mark Wulff Carstensen and Sofiia Rappe for feedback on early versions of the manuscript. 

\textbf{Author contributions:} J.K.and E.H. conceived the project and wrote the article. \textbf{Competing interests:} The authors declare no competing interests.

\bibliographystyle{plainnat}
\bibliography{biblio}

\appendix

\newpage
\section{Weak independence}\label{app:weak}

In this section, we show how Definition~\ref{def:independence} can be substantially relaxed while still ensuring our results to hold. To this end, we need to introduce another bit of formalism: We assume that predictions can be compared to establish how different they are. This is the case, e.g., in IIT where predictions map to the space of maximally irreducible conceptual structures (MICS), sometimes also called the space of Q-shapes, which carries a distance function analogous to a metric~\citep{kleiner2020mathematical}. We assume that for any given prediction, one can determine which of all those predictions that don't overlap with the given one is \emph{most similar} to the latter, or equivalently which is \emph{least different}. We calls this a \emph{minimally differing prediction} and use it to induce a notion of \emph{minimally differing data sets} below. Uniqueness is not required.

Let an arbitrary data set $o \in \O$ be given. The minimal information assumption from Section~\ref{sec:minimallyinformative} ensures that there is at least one data set $o'$ such that Equation~\eqref{eq:minimalinformation} holds. For what follows, let $o^\bot$ denote the set of all data sets which satisfy Equation~\eqref{eq:minimalinformation} with respect to $o$,
\begin{equation}
    o^\bot := \{ \, o' \in \O \ | \ \pred(\bar o) \cap \pred(\bar o') = \emptyset \ \} \:.
\end{equation}
Thus $o^\bot$ contains all data sets whose prediction completely differs from the prediction of $o$.

\begin{Def} We denote by $\min(o)$ those data sets in $o^\bot$ whose prediction is least different from the prediction of $o$.
\end{Def}

\noindent In many cases $\min(o)$ will only contain one data set, but here we treat the general case where this is not so.
We emphasize that the minimal information assumption guarantees that $\min(o)$ exists. We can now specify a much weaker version of Definition~\ref{def:independence}.

\begin{Def}\label{def:weakindependence}
\emph{Inference and prediction data are independent} 
if for any $o \in \O$ and $o' \in \min(o)$, there is a variation
\begin{equation}\label{eq:def-indep2}
    \nu: P \rightarrow P
\end{equation}
such that $o_i \in \obs(p)$, $o_i' \in \obs(\nu(p))$ but $o_r \in \obs(p)$ and $o_r \in \obs(\nu(p))$
for some $p \in P$.
\end{Def}

\noindent The difference between Definition~\ref{def:weakindependence} and Definition~\ref{def:independence} is that for a given $o \in \O$, 
the latter requires the transformation $\nu$ to exist for any $o' \in \O$, wheres the former only requires it to exist for minimally different data sets $o' \in \min(o)$.
The corresponding proposition is the following.

\begin{Prp}\label{prp:univexistsweak}
If inference and prediction data are weakly independent, universal substitutions exist.
\end{Prp}

\Proof
To show that a universal substitution exists, we need to show that for every $o \in \O$, an $o_r$-substitution exists (Definition~\ref{def:orsubst}). Thus assume that an arbitrary $o \in \O$ is given and pick an $o' \in \min(o)$. As before, we denote the prediction content of $o$ and $o'$ by $o_i$ and $o_i'$, respectively, and the inference content of $o$ by $o_r$.

Since inference and prediction data are weakly independent, there exists a $p \in P$ as well as a $\nu: P \rightarrow P$ such that $o_i \in \obs(p)$, $o_i' \in \obs(\nu(p))$, $o_r \in \obs(p)$ and $o_r \in \obs(\nu(p))$. By Definition~\eqref{eq:closureo}, the first two of these four conditions imply that $\obs(p) \subset \bar o$ and $\obs(\nu(p)) \subset \bar o'$. Since $o'$ is in particular an element of $o^\bot$, Equation~\eqref{eq:minimalinformation} applies and allows us to conclude that 
\[
\pred(\obs(p)) \cap \pred(\obs(\nu(p)) = \emptyset \:.
\]
Via Equation~\eqref{eq:Defor}, the latter two of the four conditions imply that $p \in P_{o_r}$ and $\nu(p) \in P_{o_r}$. Thus we may restrict $\nu$ to $P_{o_r}$ to obtain a map
\[
S: P_{o_r} \rightarrow P_{o_r} \:,
\]
which in light of the first part of this proof exhibits at least one $p \in P_{o_r}$ which satisfies~\eqref{eq:Subst}. Thus we have shown that an $o_r$-substitution exists. Since $o$ was arbitrary, it follows that a universal substitution exists.
\QED

The following theorem shows that Definition~\ref{def:weakindependence} is sufficient to establish the claim of Theorem~\ref{mainthm}.

\begin{Thm}\label{mainthm3}
If inference and prediction data are weakly independent, either every single inference operation is wrong or the theory under consideration is already falsified.
\end{Thm}

\Proof
The theorem follows by combining Proposition~\ref{prp:univexistsweak} and Proposition~\ref{prp:theoryfalse}.
\QED

\section{Inverse predictions}\label{app:inverse}

When defining falsification, we have considered predictions that take as input data about the physical configuration of a system and yield as output a state of consciousness. An alternative would be to consider the inverse procedure: a prediction which takes as input a reported stated of consciousness and yields as output some constraint on the physical configuration of the system that is having the conscious experience. In this section, we discuss the second case in detail.

\begin{figure}[h]
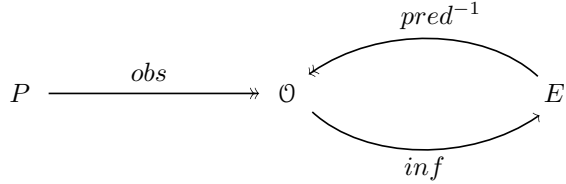

\centering
\begingroup%
  \makeatletter%
  \providecommand\color[2][]{%
    \errmessage{(Inkscape) Color is used for the text in Inkscape, but the package 'color.sty' is not loaded}%
    \renewcommand\color[2][]{}%
  }%
  \providecommand\transparent[1]{%
    \errmessage{(Inkscape) Transparency is used (non-zero) for the text in Inkscape, but the package 'transparent.sty' is not loaded}%
    \renewcommand\transparent[1]{}%
  }%
  \providecommand\rotatebox[2]{#2}%
  \newcommand*\fsize{\dimexpr\f@size pt\relax}%
  \newcommand*\lineheight[1]{\fontsize{\fsize}{#1\fsize}\selectfont}%
  \ifx\svgwidth\undefined%
    \setlength{\unitlength}{225.66017272bp}%
    \ifx\svgscale\undefined%
      \relax%
    \else%
      \setlength{\unitlength}{\unitlength * \real{\svgscale}}%
    \fi%
  \else%
    \setlength{\unitlength}{\svgwidth}%
  \fi%
  \global\let\svgwidth\undefined%
  \global\let\svgscale\undefined%
  \makeatother%
  \begin{picture}(1,0.2972074)%
    \lineheight{1}%
    \setlength\tabcolsep{0pt}%
    \put(-0.00514982,0.12986952){\color[rgb]{0,0,0}\makebox(0,0)[lt]{\lineheight{1.25}\smash{\begin{tabular}[t]{l}$P$\end{tabular}}}}%
    \put(0.44304665,0.12991773){\color[rgb]{0,0,0}\makebox(0,0)[lt]{\lineheight{1.25}\smash{\begin{tabular}[t]{l}$\O$\end{tabular}}}}%
    \put(0.88691598,0.12980107){\color[rgb]{0,0,0}\makebox(0,0)[lt]{\lineheight{1.25}\smash{\begin{tabular}[t]{l}$E$\end{tabular}}}}%
    \put(0,0){\includegraphics[width=\unitlength,page=1]{Inverse.pdf}}%
    \put(0.19737705,0.16534832){\color[rgb]{0,0,0}\makebox(0,0)[lt]{\lineheight{1.25}\smash{\begin{tabular}[t]{l}$\obs$\end{tabular}}}}%
    \put(0,0){\includegraphics[width=\unitlength,page=2]{Inverse.pdf}}%
    \put(0.64847447,0.26017193){\color[rgb]{0,0,0}\makebox(0,0)[lt]{\lineheight{1.25}\smash{\begin{tabular}[t]{l}$\pred^{-1}$\end{tabular}}}}%
    \put(0,0){\includegraphics[width=\unitlength,page=3]{Inverse.pdf}}%
    \put(0.65396182,0.00716432){\color[rgb]{0,0,0}\makebox(0,0)[lt]{\lineheight{1.25}\smash{\begin{tabular}[t]{l}$\inf$\end{tabular}}}}%
  \end{picture}%
\endgroup%

\caption{\small
The case of an inverse prediction. Rather than comparing the inferred and predicted state of consciousness, one predicts the physical configuration of a system based on the system's report and compares this with measurement results.}
\label{fig:Inverse}
\end{figure}

As before, we assume that some data set $o$ has been measured in an experimental trail, which contains both the inference data $o_r$ (which includes report and behavioural indicators of consciousness used in the experiment under consideration) as well as some data $o_i$ that provides information about the physical configuration of the system under investigation. For simplicity, we will also call this \emph{prediction data} here. Also as before, we take into account that the state of consciousness of the system has to be inferred from $o_r$, and again denote this inference procedure by $\inf$.

The theory under consideration provides a correspondence $\pred: \O \twoheadrightarrow E $ which describes the process of predicting states of consciousness mentioned above. If we ask which physical configurations are compatible with a given state $e$ of consciousness, this is simply the preimage $\pred^{-1}(e)$ of $e$ under $\pred$, defined as
\begin{equation}\label{eq:corrinverse}
\pred^{-1}(e) = \{ \, o \in \O \, | \, e \in \pred(o)  \} \:.
\end{equation}
Accordingly, the class of all prediction data which is compatible with the inferred experience $\inf(o)$ is
\begin{equation}
    \pred^{-1}\big( \inf(o) \big) \:,
\end{equation}
depicted in Figure~\ref{fig:Inverse}, and a falsification occurs in case the the observed $o$ has a prediction content $o_i$ which is not in this set. Referring to the previous definition of falsification as \emph{type-1} (Definition~\ref{def:falsi}), we define this new form of falsification as \emph{type-2}.

\begin{Def}\label{def:falsi2}
There is a \emph{type-2 falsification} at $o \in \O$ if we have
\begin{equation}\label{eq:Falsification2}
o \not \in \pred^{-1}\big(\inf(o)\big) \:.
\end{equation}
\end{Def}

In terms of the notion introduced in Section~\ref{sec:Summary}, Equation~\eqref{eq:Falsification2} could equivalently be written as $o_i \not \in \pred^{-1}\big( \inf(o_r) \big)_i$. The following lemma shows that there is a type-2 falsification if and only if there is a type-1 falsification. Hence all of our previous results apply as well to type-2 falsifications.

\begin{Lemma}
There is a type-2 falsification at~$o$ if and only if there is a type-1 falsification at~$o$.
\end{Lemma}

\Proof
Equation~\eqref{eq:corrinverse} implies that $o \not \in \pred^{-1}(e)$ if and only if $e \not \in \pred(o)$.
Applied to $e = \inf(o)$, this implies:
\[
o \not \in \pred^{-1}( \inf(o) ) \quad\text{ if and only if }\quad \inf(o) \not \in \pred(o) \:.
\]
The former is the definition of a type-2 falsification. The latter is Equation~\eqref{eq:Falsification} in the definition of a type-1 falsification. Hence the claim follows.
\QED

\end{document}